\documentclass[letter,preprint,preprintnumbers,nofootinbib]{revtex4-1}
\usepackage{amsfonts,amssymb,amsmath,color,graphicx,subfig}

\begin{document}

\title{Attractors, Universality and Inflation}

\author{Sean Downes, Bhaskar Dutta, Kuver Sinha}

\affiliation{Department of Physics and Astronomy, Texas A\&M University, College Station, TX 77843-4242, USA}

\begin{abstract}
Studies of the initial conditions for inflation have conflicting predictions from exponential suppression to inevitability. At the level of phase space, this conflict arises from the competing intuitions of CPT invariance and thermodynamics. After reviewing this conflict, we enlarge the ensemble of possible universes beyond phase space to include scalar potential data. We show how this leads to an important contribution from inflection point inflation, enhancing the likelihood of inflation to a power law: $1/N_e^3$. In the process, we emphasize the attractor dynamics of the gravity-scalar system and the existence of universality classes from inflection point inflation. Finally, we comment on the predictivity of inflation in light of these results.

\end{abstract}

\maketitle

\section{Introduction}
The original theory of big bang cosmology had a number of problems. Paramount, perhaps, was that the likelihood of our own universe seemed exponentially suppressed. Thirteen billion years was simply too old to explain our lack of curvature and assortment of celestial neighbors. The so-called age problem of cosmology was an apparent need to fine-tune the energy density of the universe to one part in $10^{60}$ \cite{peacock}. Had things not been so fine-tuned, our universe would have recollapsed or been stretched too wide to admit our own existence, and done so billions of years ago. Inflation proposes to solve this problem - and a host of others - white simultaneously giving measurable predictions that seem to agree with precision experiments \cite{Komatsu:2010fb}.

Without a full theory of quantum gravity, however, we do not have a solid understanding of how spacetime, and the physical laws it bears, emerges from the Planck scale. At first glance, inflation \cite{Linde:1981mu,Guth:1980zm,Komatsu:2010fb} appears to punt the issue of initial conditions further toward the Planck era. Not so. Despite being a CPT invariant theory, inflation is equipped with attractor dynamics \cite{Salopek:1990jq,1994PhRvD..50.7222L}. It is insensitive to the initial conditions, allowing for inflation over much of phase space. ``Much'' has come to mean different things to different people, and is the principal focus of our work.

The attracting feature of inflation has generated a great deal of work on initial conditions. In particular, it has generated some contrasting opinions. The apparent inevitability of inflation, taken with the Landscape paradigm of String Theory, has led to a deep theoretical consequence: the multiverse and the revival of the anthropic argument. In light of these discussions, Gibbons and Turok revisited the issue of initial conditions \cite{Gibbons:2006pa}.

By augmenting an earlier work of Gibbons, Hawking and Stewart \cite{Gibbons:1986xk}, they used the Hamiltonian methods to construct a measure of the phase space of a gravity-scalar system. The only inputs to this measure are the potential and a choice of time-like hypersurface to count trajectories. They choose a late-time hypersurface - which rendered the potential irrelevant - and concluded that the likelihood of having $N_e$ e-foldings of inflation in some universe $U$ is
$$\mathcal{P}(N_e|U) \propto e^{-3N_e}.$$
Since low e-foldings - and a considerably smaller universe - seems favored by the measure of \cite{Gibbons:2006pa}, their proposed problem resembles the age problem. Closer inspection, however, suggests the problem lies with non-adiabatic expansion. This is a classical statement about the metric\footnote{In particular, this is not a quantum statement about a scalar field. During slow roll inflation the scalar field have a small mass and is approximately conformal. Nonlinearities are effectively absent and its \textit{perturbations} are also said to be adiabatic, but in the quantum context and for different reasons.}. The findings of \cite{Gibbons:2006pa} suggest that cooling, \textit{adiabatic} expansion of the universe is strongly favored. This means no accelerated expansion, no density perturbations and none of the theoretical solutions of inflation should ever be expected in a generic universe. In such a case inflation has no trigger; any observed inflation might need to appeal to the anthropic princple for explaination.

A late time application of the cosmological measure seems naturally motivated. The classical gravity-scalar system is CPT invariant, and the relevant dynamics for a single scalar field are not chaotic. Since the UV physics is not accessable, it would seem to make sense to simply apply the Cauchy data at a late time, where we have observed around sixty e-foldings of inflation.

Closer study reveals a subtle juxtaposition of CPT invariant field equations with non-adiabatic solutions. Arguing for the scarcity of inflationary trajectories by tracing them backwards in time is something akin to reconstructing a wine glass from a pile of shards. As pointed out in \cite{kofman,Linde:2007fr}, running an attractor solution - like inflation - backwards in time gives rise to a dynamical repuslor. Paradoxically, the gravity-scalar system appears to respect both time reversal invariance and the second law of thermodynamics. 

To conciliate the thermodynamic laws and CPT invariance, one observes that a highly ordered state is universally unlikely. Just as it is rare for a glass to self-assemble once broken, so too is it rare for glass to exist in the first place! This appears to be the germ of the argument in \cite{Gibbons:2006pa}. A peculiar feature about this multiverse measure is that the likelihood of inflation seems to depend on which time-like hypersurface you pose the question. This ambiguity has long been known \cite{Hawking:1987bi}. In chaotic inflation, this means counting inflationary trajectories either just after universe was born, or just before reheating. The former gives preference to inflation. The latter suggests it is exponentially suppressed, but has the advantage that it applies Cauchy data near to what we actually observe. It would appear that the likelihood of exponential expansion depends critically on the value of the Hubble parameter. The paradox at hand comes from the conflation of irreversible processes and attractor dynamics; in the context of cosmology this may be quite wrong!

The aim of this work is to deconflate these two ideas. Specifically, we show how a CPT invariant theory can scramble a wide range of initial conditions and give rise to inflation. In a sense, we demonstrate how some ``cosmological'' glasses can easily self-assemble after breaking. Our vessel to illustrate this point will be inflection point inflation. The scrambling agents on phase space will be the singularities associated with degenerate critical points of the scalar potential. Importantly, this scrambling remains intact after resolving the singularities.

Counter to one's intuition -- and many claims to the contrary -- inflection point inflation does not require an elaborate fine tuning of initial conditions. Simple, but careful application of the Friedmann constraint to the scalar field equation is all that is needed to observe this. Indeed, in \cite{Itzhaki:2008hs} Itzhaki and Kovetz found the existence of a ``phase transition'' between chaotic and inflection point inflation. Roughly speaking, this transition occurs when the height of the potential at the inflection point assumes a minimum value. Specifically, they demonstrated how such degenerate critical points will attract virtually all trajectories. Part of our work will be to generalize their result and to understand this transition physically and generalize the results towards a complete understanding of these ``efficient'' attractor points.  

Specifically, we enlarge the ensemble of universes discussed in \cite{Gibbons:1986xk} to include variations of the scalar potential. In an effort to be as model independent as possible, we focus on potentials that are monotonically decreasing during the period of inflation\footnote{With care, this analysis can be extended to include potentials with local minima, which may lead to other kinematic or quantum features. We will not consider those effects here. For the insistent reader, one many assume that any such effects happened before our analysis begins.}. Taking the approach of \cite{Downes:2011gi}, we then analyze these potentials locally using V.I. Arnold's theory of singularities \cite{arnold}. Physically, one may interpret this as allowing the couplings in the scalar potential from the effective field theory to vary in the ensemble. The results of this analysis suggest that contributions from inflection points give a power law likelihood for the number of e-foldings in any given universe.

While it seems that nothing short of a full understanding of UV complete physics can mandate a specific choice of time-like hypersurface in \cite{Gibbons:2006pa}, our analysis holds well below the Planck scale and hopefully strengthens the case for a predictive power of inflation. At least at the classical level, we will show that the emergence of inflection points resolve this ambiguity from a continuously varying answer to one exponentionally close to a yes-no question.

The remainder of the paper is organized as follows: In section \ref{sectiontwo} we give a brief review of inflation with a special emphasis on attractor dynamics. In the process we extend the analysis of Kofman and Linde \cite{Linde:2007fr,kofman}. Next, in section \ref{sectionthree} we discuss inflection point inflation, introduce and extend the work of Itzhaki and Kovetz \cite{Itzhaki:2008hs}. In section \ref{sectionfour} we apply these ideas to the likelihood of inflation. In section \ref{sectionfive} we draw our conclusions. We have also included an appendix to frame our statistical discussion and make precise what we mean by the likelihood of inflation.

\section{Field Equations}\label{sectiontwo}
We begin our analysis by reviewing the set up for inflation in the classical gravity-scalar system. By parametrizing time with the scale factor, the attractor dynamics manifests itself the field equation for the inflaton. We then explore the attractor-replusor dynamics with an eye towards inflation. These attractor dynamics becomes crucial when discussing inflection point inflation in the following section.
\subsection{Derivation}
The field equation for the scalar field in the FLRW-background is
\begin{equation}\label{KG}\ddot{\phi}+3H\dot{\phi}+\frac{\partial V}{\partial \phi} = 0.\end{equation}
Dots represent the usual time derivative. The Hubble parameter, $H$ is $\dot{a}/a$. The dynamics scale factor $a$ is governed by the Friedmann constraint,
\begin{equation}\label{easyfried}3M_P^2H^2 = \frac{1}{2}\dot{\phi}^2 + V,\end{equation}
and the evolution equation,
\begin{equation}\label{easyevo}\dot{H} = -\frac{1}{2M_P^2}\dot{\phi}^2.\end{equation}
$H$ appears as part of the covariant derivative (with respect to time) of $\dot{\phi}$. The evolution and covariant Klein Gorden equations are a pair of coupled nonlinear differential equations for $a$ and $\phi$, subjected to \eqref{easyfried}. Our aim is to understand the solutions to these equations in a manner that makes their attractor properties manifest.

Our first step in solving these equations is to separate them. To that end, we define a new variable $N$,
$$N=\log\left(a(t)/a_0\right)$$
This is the parameterization familiar from a Hamilton-Jacobi analysis of inflation, but also from studies relevant for our analysis\cite{1994PhRvD..50.7222L,Linde:2007jn,Allahverdi:2008bt}. Notice that
$$ \frac{dN}{dt} = H.$$
Effectively, this parametrizes time with the scale factor $a$. Note that parametrization is a bijection only if $V$ is nonzero. While important for studying reheating physics near the global minimum of V, our work focuses on inflation and will never need to consider this case.

An important consequence of this parametrization comes from the chain rule:
$$\dot{\phi} = H\phi^{\prime}.$$
Here $\phi^{\prime} = d\phi/dN$. The Friedmann equation becomes,
\begin{equation}\label{newfried}H^2 = \frac{1}{3M_{P}^2}\big(\frac{1}{2}\phi^{\prime 2}H^2 + V\big).\end{equation}
So long as $|\phi^{\prime}|<\sqrt{6}M_P$, we have a new expression for the Hubble parameter,
\begin{equation}\label{hubble}H^2 = \frac{1}{3M_{P}^2}\frac{V}{1-\frac{1}{6 M_{P}^2}\phi^{\prime 2}}.\end{equation}
One may interpret the singular behavior of the system at $|\phi^{\prime}|=\sqrt{6}M_{P}$ is an artifact of our parametrization, but it plays an important role in the analysis that follows.

 We are now close to isolating an equation for $\phi$ only. Computing $\ddot{\phi}$,
$$\ddot{\phi} = \dot{H}\phi^{\prime} + H^2\phi^{\prime\prime},$$
we find
$$\ddot{\phi} = H^2\big(\phi^{\prime\prime}-\frac{\phi^{\prime 3}}{2M_{P}^2}\big).$$
Therefore,
$$H^2\big(\phi^{\prime\prime} -\frac{\phi^{\prime 3}}{2M_{P}^2} + 3\phi^{\prime}\big) + \frac{\partial V}{\partial \phi}=0.$$
A final simplification gives the ``separated'' nonlinear differential equation for the dynamics of $\phi$:
\begin{equation}\label{eom}\phi^{\prime\prime} +3\big(1-\frac{\phi^{\prime 2}}{6M_{P}^2}\big)\Big[\phi^{\prime} + M_P^2\big(\frac{\partial}{\partial \phi}\log V\big)\Big]=0.\end{equation}
The attractor dynamics can be read off simply from this form, as we now discuss.
\subsection{Asymptotic Solutions}

By analyzing the structure of \eqref{eom}, we can determine the behavior of its solutions analytically. Specifically, we shall focus on the three ``singular trajectories'',
$$\phi^{\prime} = \pm \sqrt{6}M_{P}\quad \mathrm{and}\quad \phi^{\prime} = -M_{P}^2\frac{\partial \log V(\phi)}{\partial\phi}.$$
Not only can we extract physically important information - like the number of e-foldings - analytically, but this approach will better facilitate the discussion in later sections of inflection point inflation and how it affects the measure problem in cosmology.
\subsubsection{Small $\phi^\prime$}
Let $\phi_{\star}$  be the singular ``trajectory'' which solves
\begin{equation}\label{slowroll}\phi^{\prime} + M_P^2\big(\frac{\partial}{\partial \phi}\log V\big) = 0.\end{equation}
Owing to the nontrivial $\phi$ dependence of $V$, this will not generically solve the full equations of motion. Indeed, to be a true solution, $\phi_{\star}^{\prime\prime}$ must vanish. To that end, consider
$$\phi^{\prime\prime} = -M_{P}^2\Big[\frac{1}{V}\frac{\partial^2 V}{\partial \phi^2} - \Big(\frac{1}{V}\frac{\partial V}{\partial \phi}\Big)^2\Big]\phi^{\prime}. $$
Thus, $\phi_{\star}$ is an approximate solution to the full field equation, \eqref{eom}, so long as the so-called \textit{slow roll conditions},
\begin{equation}\label{slowrollcond}M_{P}^2\Big(\frac{1}{V}\frac{\partial V}{\partial \phi}\Big)^2\ll 1,\quad M_{P}^2\frac{1}{V}\frac{\partial^2 V}{\partial \phi^2}\ll 1,\end{equation}
are satisfied. In short, $\phi_{\star}$ is the trajectory undergoing slow-roll inflation.
As a concrete example, consider a monomial potential 
$$V = \phi^{2n}.$$
The singular or slow-roll trajectory is given by
$$\phi^{\prime} = -\frac{2n}{\phi}M_{P}^2.$$
Notice that
$$\phi^{\prime\prime} = \frac{2n}{\phi^2}M_{P}^2.$$
Thus, this solution is approximately valid so long as
$$\phi \gg M_{P}.$$
As we will see in the next subsection, this and the other singular trajectories, $\phi^{\prime}=\pm\sqrt{6} M_{P}$ strongly influence the behavior of generic trajectories. This influence is strong enough to extract information about virtually any trajectory analytically. Moreover, this formulation of the slow-roll trajectory will be useful in discussing inflection point inflation below. Henceforth we set $M_{P}=1$.

\subsubsection{Large $\phi^\prime$}
There are a pair na\"{i}ve solutions to \eqref{eom}, $\phi^{\prime}=\pm \sqrt{6}$. As a singular solution, $\phi$ would grow linearly with $N$ perpetually, or at least so long as the relation between time and $N$ is well-defined. From \eqref{newfried} we see that that such a solution occurs only when $V$ vanishes; $|\phi^{\prime}|$ is bounded below by $\sqrt{6}$ by the Friedmann equation. These solutions - like finite energy solutions near to it - are not inflating. $\phi$ behaves as radiation and the scale factor obeys a power law, $a\sim 1/\sqrt{t}$. Nevertheless, this singular solution has an important residual impact on the physics - even for very large $\phi$. We now discuss these effects. 

\begin{figure}[h]
  \centering
    \includegraphics[width=0.8\textwidth]{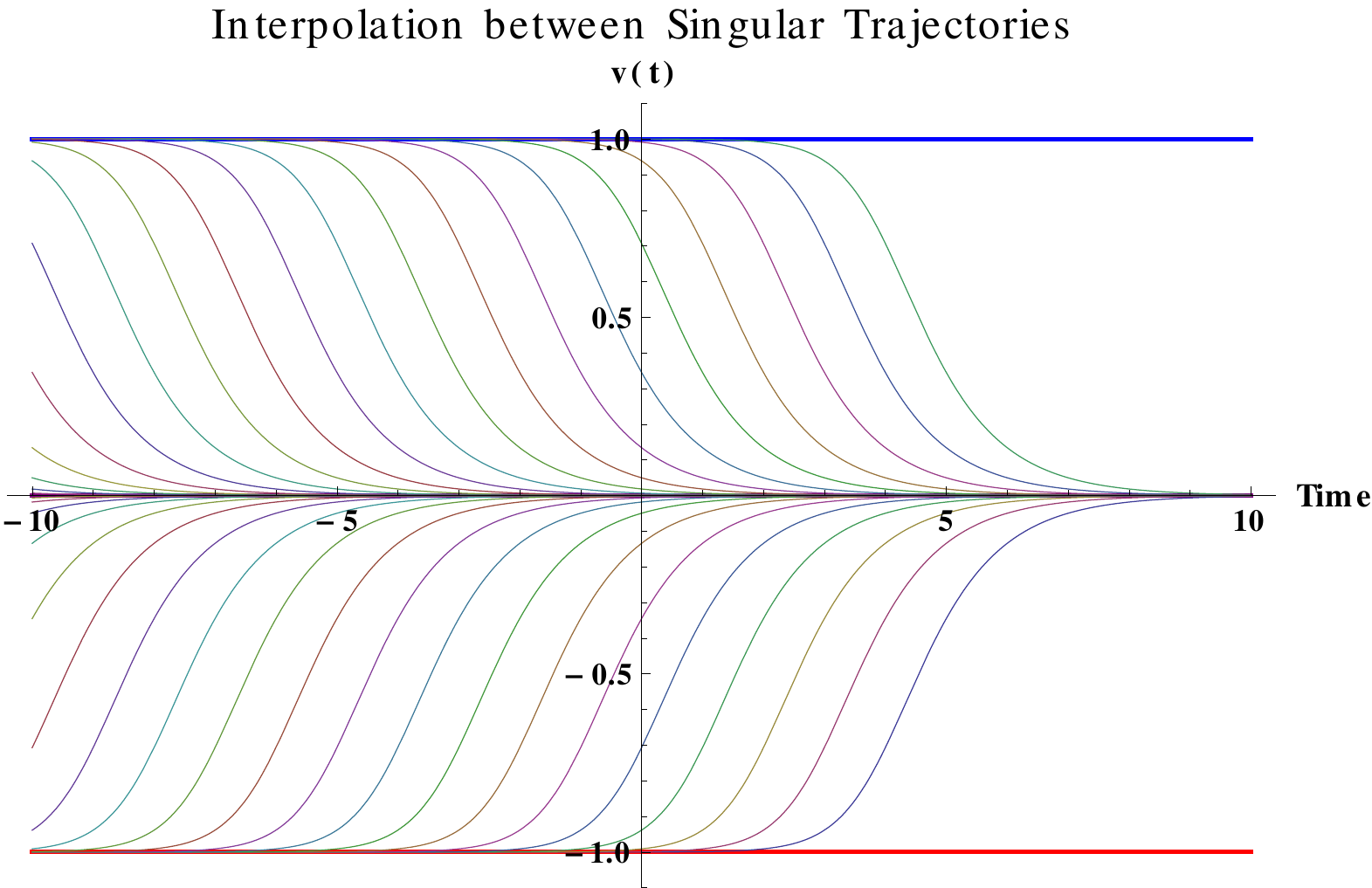}
  \caption{Solution of the differential equation $\dot{v} = v(v-1)(v+1)$ as a proxy for Eqn.~\eqref{epsilon}, where we make the reduction to a first order differential equation explicit. Note the three singular trajectories at $v=\pm 1,0$, between which all other solution interpolate.}

\label{triples}
\end{figure}

These singular solutions are repulsor trajectories in phase space. To see that, let us perturb around them.
$$\phi^{\prime} = \pm(\sqrt{6}-\epsilon).$$
For definiteness, we choose the minus sign. The reparameterized Klein Gordon equation \eqref{eom} gives an equation for $\epsilon$.
\begin{equation}\epsilon^{\prime} = -3\epsilon\big(\frac{2}{\sqrt{6}} + \frac{\epsilon}{6}\big)\big(-\sqrt{6}+\epsilon + \frac{\partial \log V}{\partial\phi}\big).\end{equation}
We simplify this by introducing a scaling index for $V$, $\nu$.
$$\nu = \frac{\partial \log V}{\partial\log\phi},$$
which yields
\begin{equation}\label{epsilon}\epsilon^{\prime} = \frac{1}{2}\epsilon(\epsilon-2\sqrt{6})(\epsilon-\sqrt{6}+\nu/\phi).\end{equation}
Before dealing with the $\phi$ dependence in the third factor of the-right hand-side, let us investigate the structure of this differential equation. To that end, suppose $\phi$ is fixed.
There are three singular solutions to \eqref{epsilon},
\begin{equation}\epsilon = 0,\quad \epsilon = \sqrt{6}-\nu/\phi,\quad \mathrm{and}\; \epsilon = 2\sqrt{6}.\end{equation}
If these three solutions were indeed constant, then study of this differential equation reduces to two kinds of solutions: the aforementioned three ``vacua'', and ``solitons'' which interpolate between them. Qualitatively, the greatest and least of the three critical solutions are repulsors. All solutions near them are drawn away exponentially and asymptotically approach the middle solution; it is an attractor\footnote{This applies to solutions which take values \textit{between} the two repulsor trajectories. Solutions above or below these trajectories are also repelled; they are divergent.}. Thus, if
$$0 <\sqrt{6}-\nu/\phi < 2\sqrt{6},$$
$\phi^{\prime}$ is repelled from $\sqrt{6}$ and approaches $\nu/\phi$. In other words, the slow roll solution \eqref{slowroll} is an attractor. This behavior is sketched in Fig.~\ref{triples}. 

We now discuss the implications of this for $\phi$ with an eye towards inflation. Happily, this attractor/repulsor behavior exists regardless of the $\phi$ dependence of $\nu/\phi$. The primary effect of varying $\phi$ is to exchange the roles of a repulsor and an attractor. That is, when $\sqrt{6}<\nu/\phi$, $\phi^{\prime}$ is repelled from the slow roll trajectory and begins to approach $\sqrt{6}$. For a quadratic potential, for instance, this occurs when $\phi\sim \sqrt{2/3}$. Thus, the critical solution $\phi^{\prime} = -\sqrt{6}$, now an attractor, ends inflation by pulling $\phi$ away from slow roll.

The number of e-foldings is then
\begin{equation}\label{fullne}N_e \approx \sqrt{6}(\phi_{i}-\phi_t) + \int_{\phi_{\rm end}}^{\phi_t} d\phi\;\frac{1}{\partial\log V/\partial \phi},\end{equation}
where $\phi_i$ is where the initial field value, transition takes place at $\phi_t$ and  $\phi_{\rm end} \sim \nu/\sqrt{6}$. We shall see explicit applications of this formula in section \ref{sectionfour}.

\begin{figure}[h]
  \centering
    \includegraphics[width=0.8\textwidth]{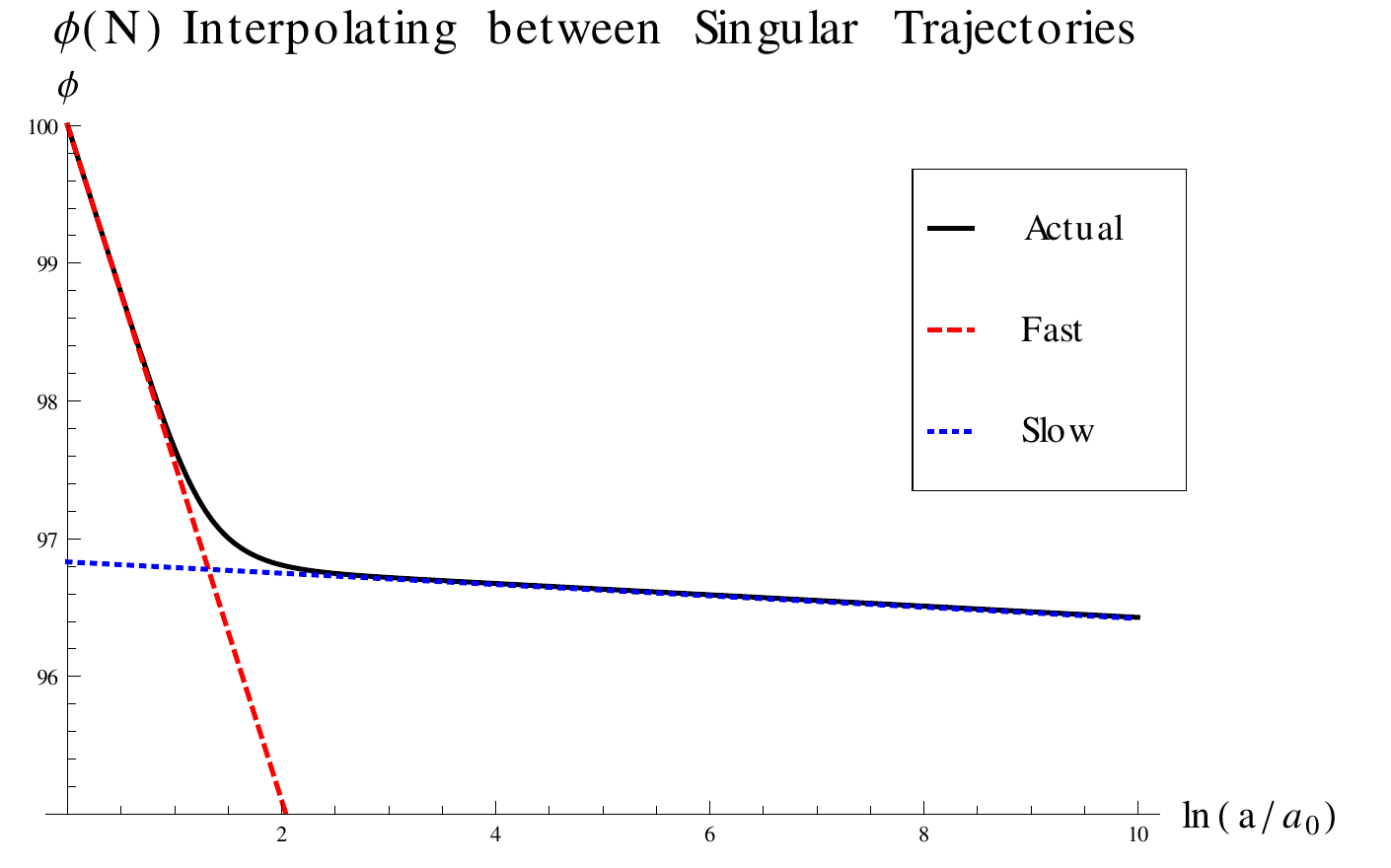}
  \caption{Trajectory that starts near the fast roll trajectory $(\phi,\phi^{\prime})= (100,-\sqrt{5.99})$ that ``decays'' to the slow roll trajectory, $\phi^{\prime}\sim- \partial \log V/\partial \phi$.}
\label{fig:trajectory}
\end{figure}

Morally, studying singular trajectories of \eqref{eom} is akin to the study of solitons. The famous attractor behavior of inflation is now directly understood from \eqref{eom}, as is the way inflation ends. Indeed, for a generic set of initial conditions, one finds three distinct regimes: a relatively short linear (in $N$) where $\phi^{\prime} \sim \sqrt{6}$, a quick transition\footnote{It is possible that the field will overshoot the minimum of the potential in the linear phase. In this case, the transition regime will typically correspond to a turnaround. $\phi$ then rapidly approaches the slow roll phase on the way back down.}, and a longer period of slow roll inflation. This is demonstrated numerically integrating the field equations in Fig.~\ref{fig:trajectory}.

The considerations apply to a generic, smooth potential. To get enough e-foldings of inflation to match observation, however, one typically needs initial conditions where the VEV of $\phi$ is very large. Typically $\big<\phi\big> \gg M_p$. This is the chaotic inflation scenario. Alternatively, if the first and second derivatives of the potential (nearly) vanish together at a point, arbitrarily large amounts of inflation may be possible. This is inflection point inflation, which we now discuss in detail.

\section{Inflection Points}\label{sectionthree}

Having discussed the attractor dynamics of inflation, we open this section by sketching the physics of inflection point inflation. In particular, we examine the dissipative behavior of the ``derived'' inflaton field equation, \eqref{eom}. This effective dissipation gives rise to a ``phase transition'' between chaotic and inflection point inflation pointed out in \cite{Itzhaki:2008hs}. This is another manifestation of attractor dynamics, first studied in this context in \cite{Allahverdi:2008bt}. We describe how it serves to scramble almost all initial conditions and gives rise to an additional potential-dependent number of inflationary e-foldings. In other words, the attractor dynamics frequently provides appropriate initial conditions for inflection point inflation. We then describe specifically when it fails. When then discuss how these effects depend on the parameter (coupling) space of the scalar potential. In the following section, we apply this analysis to the ensemble of possible universes to determine the likelihood of inflation.

\subsection{Inflection Point Inflation}
In our discussions above, the slow roll conditions \eqref{slowrollcond} were satisfied using large field VEVs. This is the well-studied paradigm of chaotic inflation. There is another way to satisfy \eqref{slowrollcond}, namely a point in field space where
$$V^{\prime}\approx 0,\quad V^{\prime\prime}\approx 0.$$
That is, a value of $\phi$ were $V$ has a degenerate critical point. With the appropriate initial conditions, one can integrate \eqref{slowroll} to obtain the number of e-foldings,
$$N_{e} = \int d\phi \frac{V}{\partial V/\partial \phi}.$$
When $V^{\prime}$ is vanishing small, $N_e$ diverges. At first glance, one may view inflation in such a model as ``accidental''. Fortunately, attractor dynamics amplifies the effectiveness of this approach by setting up the initial conditions. Discussion of these initial conditions is the main focus this section. First, however, we compute the number of e-foldings. 

As discussed below, the parameter space of a generic potential $V$ includes domain walls of co-dimension one. On these domain walls, $V$ has a degenerate critical point. All but a set of measure zero\footnote{The set points in parameter sapce that correspond to $n$-fold degenerate critical points typically form a hypersurface of co-dimension $n-1$.} are doubly degenerate, giving a cubic behavior suitable for inflation. Generically, slight deformations of the potential will break the degeneracy. By expanding around the point of inflection $\phi = \varphi-\alpha$, this can be accounted for by a linear term, $\lambda_1$.
\begin{equation}\label{need} V = V_0\big(1+\lambda_1 \varphi + \frac{\lambda_3}{3} \varphi^3 + \mathcal{O}(\varphi^4)\big).\end{equation}
In other words, inflation can be expected if $\lambda_1 \ll \lambda_3$. As discussed in \cite{Downes:2011gi}, the overall scaling $V_0$ and the cubic co-efficient $\lambda_3$ depend strongly on the parameters along the domain wall, and negligibly on ``morsification'' parameter $\lambda_1$.
The number of e-foldings is then given by
$$N_e = \int_{0}^{\varphi_{\rm END}} \frac{V_0}{\lambda_3 \varphi + \lambda_1} + \mathcal{O}(\varphi).$$
Where $\varphi_{\rm END}$ marks the end of inflation. Computing, one finds
$$N_e= V_0\frac{\arctan\big(\phi\sqrt{\lambda_3/b}\big)}{\sqrt{\lambda_1 \lambda_3}} \Big|_{0}^{\varphi_{\rm END}}.$$
The integral is dominated by the inflection point $\varphi=0$, which to a very good approximation allows one to take the limit $\varphi_{\rm END}\rightarrow\infty$. Thus,
\begin{equation}\label{infKNEE}N_e\approx \frac{V_0\pi}{2{\sqrt{\lambda_1 \lambda_3}}}.\end{equation}
\subsection{Inflection Point Attractors}
\subsubsection{Dissipation}
By integrating out the dynamics of the metric, the scalar field $\phi(N)$ is no longer a Hamiltonian system. There is dissipation (Hubble friction), which is responsible for the attractor behavior mentioned above. Due to this ``Hubble friction'', critical points of the potential stand to be attractor points of the dynamics. For example, the (classical) field will eventually come to rest at the bottom of some local minimum of $V$.


The energy density of the scalar field is given by,
$$\rho = \frac{V}{1-\frac{1}{6}\phi^{\prime 2}}.$$
Taking a derivative we find,
$$\rho^{\prime} = \frac{\phi^{\prime}V}{\left(1-\frac{1}{6}\phi^{\prime 2}\right)^2}\Big[\frac{(1-\frac{1}{6}\phi^{\prime 2})}{V}\frac{\partial V}{\partial \phi} + \frac{\phi^{\prime\prime}}{3}\Big].$$
Applying \eqref{eom},
$$\rho^{\prime} = \frac{\phi^{\prime}V}{\left(1-\frac{1}{6}\phi^{\prime 2}\right)^2}\Big[(1-\frac{1}{6}\phi^{\prime 2})\frac{\partial \log V}{\partial \phi} - (1-\frac{1}{6}\phi^{\prime 2})(\phi^{\prime} + \frac{\partial \log V}{\partial \phi})\Big],$$
so that finally,
\begin{equation}\label{rhop}\rho^{\prime} = -\rho \phi^{\prime 2}.\end{equation}
As one would expect from \eqref{KG}, the larger $\rho$, the larger $H$, and therefore, the larger the dissipation. This brings about the possibility that enough energy may be dissipated to set up inflection point inflation. As we shall see, this possibility is realized. Indeed, it gives added physical insignt into on some recent work by Itzhaki and Kovetz.
\subsubsection{The Itzhaki and Kovetz Model}
\begin{figure}[h]
  \centering
    \includegraphics[width=0.8\textwidth]{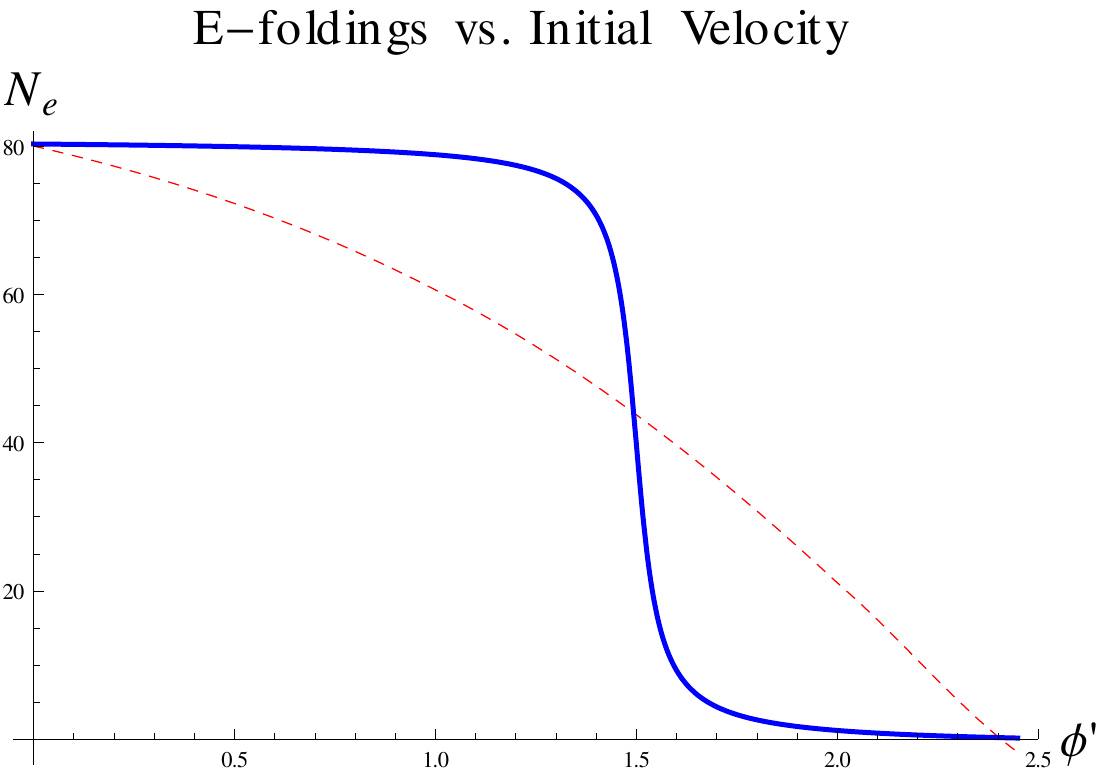}
  \caption{E-foldings versus initial Hubble velocity for the inflection point potential \eqref{A4} (solid) and a simple quadradic (dashed). Both potentials were rigged to yield precisely $N_e=80$ when started from rest to demonstrate the different behaviors.}
\label{fig:velo}
\end{figure}

In \cite{Itzhaki:2008hs}, Itzhaki and Kovetz analyzed the potential,
\begin{equation}V = \alpha \phi^3 + 1,\end{equation}
and found that below $\alpha = \alpha_C \approx 0.774$, the scalar field came to rest at the inflection point regardless of its initial position. Despite their restriction on the inital velocity, the system attracts to the inflection point for a wide range of initial velocities, as shown in Fig~\ref{fig:velo}. Above $\alpha_C$, the field slowed down near the inflection point, but eventually passed it by. They went on to demonstrate that this change had the structure of a second order phase transition. In particular, they showed the higher order corrections, $\mathcal{O}(\phi^{4})$ have no influence on their observed critical exponent. Taken with \cite{Downes:2011gi}, the results of \cite{Itzhaki:2008hs} are universal for arbitrary smooth potentials with inflection point. We comment further on this in the next subsection.

To appreciate the physics of the Itzhaki-Kovetz result let of first rewrite $V$ as
$$V = \alpha(\phi^3 + 1/\alpha).$$
As \eqref{eom} depends only on the logarithmic derivative of $V$, the overall scaling is unimportant. From this perspective, we see that the height of the potential at the inflection point, $1/\alpha$, determines the attractor behavior. With \eqref{rhop} in mind, we see that if $V$ takes a value above $1/0.774\sim1.29$, the friction will be sufficient to stop the field at the inflection point. That this is the case can indeed be verified numerically, which is how we reproduced their result for $\alpha_C$. Moreover, this logic can be applied to any potential with an inflection point.


One final comment. In \cite{Itzhaki:2008hs}, the initial conditions considered had only zero velocity. In light of section \ref{sectiontwo}, this means that $\phi$ was already close to the slow roll trajectory. If the field starts with a large velocity, and doesn't have a long enough trajectory to dissipate the energy, it will overshoot even if $\alpha>\alpha_C$. This observation will play an important role in section \ref{sectionfour}, and we will develop it further then.


\subsection{Parameter Space}
Generally, the potential of the scalar field depends on some external parameters or couplings. The physics of the overall system is sensitive to these parameters, as is familiar from the study of spontaneous symmetry breaking. The stable configurations of a system depend on the set of local minima. As critical points of the potential pass from real to complex values, the number of local extrema changes in a predictable way. Configurations with different numbers of local minima are often called different ``phases''.

When critical points degenerate, the system is only marginally stable and is the boundary between two distinct phases. 
Thus, configurations of the potential with degenerate critical points divide different domains in parameter space. The geometry of the bounding ``domain walls'' will be relevant for discussing the likelihood of inflation below. We give a quick example of a two-parameter model, and then state the analogous, relevant features for arbitrary models. The general feature is same as the one parameter model discussed before. For a more detailed discussion associated with this two parameter model, see \cite{Downes:2011gi}.
\vskip1ex
Consider a suitably normalized quartic potential,
\begin{equation}\label{A4}V = \frac{1}{4}x^4 + \frac{1}{2}a x^2 + bx + c.\end{equation}
The cubic term in $V$ can always be eliminated by a shift, corresponding to choice of $c$. Since we are interested in the topological behavior of the parameter space, we ignore $c$ for this discussion.
The existence of degenerate critical points requires $V^{\prime}=V^{\prime\prime}=0$. This amounts of the constraint,
\begin{equation}\label{cusp}\Big(\frac{a}{3}\Big)^3 + \Big(\frac{b}{2}\Big)^2 = 0.\end{equation}

\begin{figure}[h]
  \centering
    \includegraphics[width=0.7\textwidth]{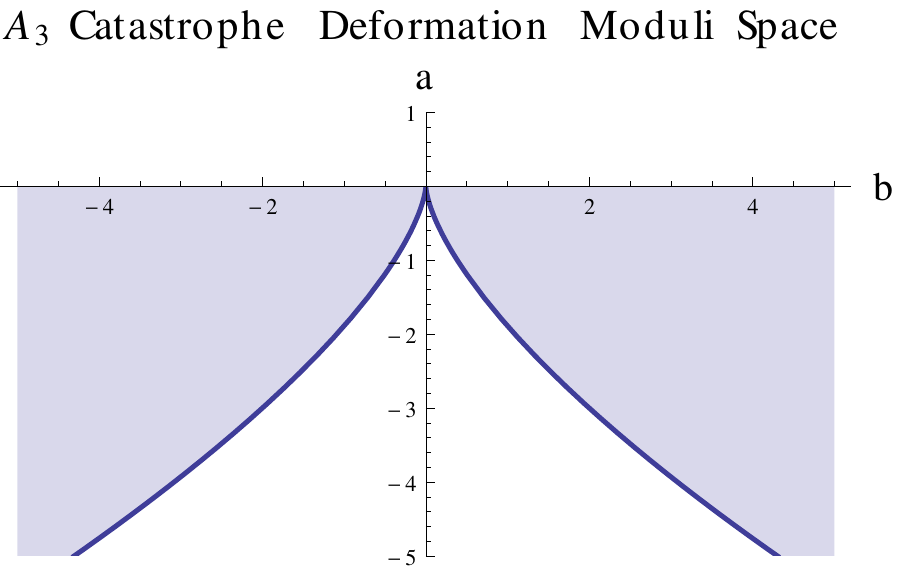}
  \caption{Parameter space of the potential in Eqn.~\ref{A4}, with the constant term ignored. The line represents a ``domain wall'' of codimension one specified by Eqn.~\ref{cusp}. The shaded region corresponds to potentials suitable for inflection point inflation.}
\label{cuspfigure}
\end{figure}

This condition divides the space of parameters $(a,b)$ in two, as shown in Fig.~\ref{cuspfigure}. The upper region has a single minima at the origin; the lower region has two distinct minima. When \eqref{cusp} is satisfied, there is a degenerate critical point suitable for inflation. Note that, for a two parameter model, the (perturbative) physics is dominated at high energies by a quartic term. For (four dimensional) $n$ parameter models, the high energy limit corresponds to an $n-2$ dimensional operator. 

For a model with an $n$-dimensional parameter space $P_N$, the condition for a doubly degenerate critical point is a co-dimension one hypersurface $\mathcal{D}$ in $P_N$. A triply degenerate critical point corresponds to a hypersurface of co-dimension two. These conditions form a complex, included entirely with $\mathcal{D}$, which maps out the phase diagram of the theory. It typically converges at (n+2)-fold degenerate critical points, corresponding to the single $n-2$ dimensional operator. We will, however, only be interested in doubly degenerate critical points \footnote{For $m>2$, the points in $P_N$ corresponding to $m$-fold degenerate critical points are a set of measure zero in $\mathcal{D}$. Assuming inflection point inflation takes place, we need only consider those with doubly degenerate critical points - cubic behavior. Of course,  $\mathcal{D}$ is itself a set of measure zero in $P_N$. Fortunately, inflation ends, and we need only consider potentials whose parameters take values \textit{near} $\mathcal{D}$. }.

\subsubsection*{Application to the Itzaki-Kovetz Model}
We end this section with a final observation. Using \eqref{cusp}, we can rewrite \eqref{A4} as
\begin{equation}V(x) = \frac{1}{4}(x-\alpha)^4 + \alpha (x-\alpha)^3 + \frac{27}{4}\alpha^4.\end{equation}
Here, $\alpha = (-a/3)^{1/2} = (b/2)^{1/3}$. There is a local minimum at $x = -2\alpha$, and we have chosen the constant term for the potential vanishes there. Careful inspection show that $\alpha$ is both the location of the inflection point and a parametrization of the relation \eqref{cusp}. Importantly, it also parametrizes the energy density at the inflection point - the scale of inflation. As discussed in \cite{Downes:2011gi}, the scale of inflation generically scales with coefficient of the cubic term. Therefore, given the Itzaki-Kovetz result, we see that such a critical $\alpha_C$ will always exist. Moreover, since any arbitrary smooth potential will be associated to a specific catastrophe germ (i.e. universality class, see \cite{Downes:2011gi}), this scaling will determine the approximate value of $\alpha_C$. For example, in the discussed above the critical value of $\alpha$ is $\alpha_C\approx 0.660$.

\section{The Likelihood of Inflation}\label{sectionfour}
In this section we first discuss the measure problem in cosmology and review some recent work related to it, particularly that of Gibbons and Turok. We then apply the results from the previous section to extend the measure over the space of possible scalar potentials. We discuss how this gives rise to a power law likelihood for inflation, in agreement the numerical results of \cite{Agarwal:2011wm}.
\subsection{Overview}
\subsubsection{Motivation}
The initial conditions for inflation has been studied a great deal \cite{Gibbons:1986xk,Gibbons:2006pa,Hollands:2002xi,Hawking:1987bi}. Particular attention has been paid to the overall likelihood of inflation. That is, the expectation value of the number of e-foldings, $\big<N_e\big>$ for an ensemble of possible universes. Tying observations to inflation, our universe must have gone through around sixty e-foldings of inflation since the perturbations observed by COBE left the horizon, but there may be many more.

The authors of \cite{Kofman:2002cj} and \cite{Gibbons:2006pa} view the attractor dynamics from constrasting perspectives. The CPT invariance of the measure is weighed against the second law of thermodynamics; each argument would be theoretically appealing if not for the other. At the core of this paradox is that FLRW spacetime manifestly violates time reversal invariance. In spite of this, ``Hubble friction'' does not dissipate energy. Classically, the Hamiltonian is always zero; this is a constraint on the system. Quantum mechanically, the inflationary universe is deficit spending: the quantum fluctuations -- variations in energy borrowed against an uncertainty in time -- normally dissapear into the vacuum. Instead, they become classical because of the exponential expansion of space. This extreme violation of the Heisenberg's original uncertainty principle is driven by the attractor dynamics. Understanding this interplay of classical and quantum physics requires us to disentangle the physical ideas from CPT invariance and thermodynamics. For this reason we focus on the Gibbons-Turok \cite{Gibbons:2006pa} approach.

\subsubsection{The Argument}
Arbitrarily large $N_e$ is possible in principle. As it happens, the analysis of Gibbons and Turok suggest that likelihood of being in a universe that has undergone $N_e$ e-foldings of inflation is $\sim e^{-3N_e}$, an exponential suppression! If $\big<N_e/60\big>$ is extremely suppressed we are left with three possibilities. First, inflation may be an incomplete theory - or wrong altogether. Second, our universe is exceedingly improbable and we are lucky to be alive. Third, our ensemble average itself is somehow incomplete. 

Without a working knowldege of physics above the Planck scale, we cannot have an answer to the first possibility. The second possibility - the anthropic solution - has been promoted lately in lieu of any apparent vacuum selection mechanism in string theory. From both of these options, inflation loses predictive power. It can model the data beautifully, but cannot explain it.

Therefore, we suggest the third option: the measure itself needs augmentation. In this section, we first review the ensemble of \cite{Gibbons:2006pa}. Next, we expand the ensemble to include the parameters (couplings) of the inflaton potential. Finally, we shall revisit the issue of selecting an appropriate time-like hypersurface to compute the likelihood of inflation.

\subsection{The Gibbons-Turok Argument}
\subsubsection{CPT versus Thermodynamics}
As we have seen, inflation is an attractor solution. If a primordial scalar field exists - and string theory suggests hundreds do - the universe should tend towards exponential inflation. In \cite{Kofman:2002cj}, the authors make a simple argument why the initial conditions for inflation should not matter. Given a sufficently shallow potential, inflation is inevitable unless the field is severely constrained to start with a VEV near a local minimum of its potential.

In \cite{Gibbons:2006pa}, Gibbons and Turok provided such a constraint. Since their techniques were CPT invariant, they performed their calculations at late times and extrapolated backwards to find the severe constraints on the initial conditions. Running an attractor solution backwards in time created a dynamical repulsor, which was the origin on the suppressed field VEV. This strongly suggests a violation of the second law of thermodynamics. But the physics is not yet clear. Before discussing this issue further, we review the cosmological measure of Gibbons and Turok .

\subsubsection{The Cosmological Measure and its Function}
The cosmological measure adopted in \cite{Gibbons:1986xk} is the analog of the density of states for an emsemble of universes. This can be interpreted as giving a sympletic structure $\omega$ on phase space. Before applying any constraints, the 2n-dimensional na\"ive phase space has a volume from $\omega^n$, which is commonly written as 
$$\omega^n \rightarrow \hat{\rho}(p,q)d^np d^nq,$$
where $\hat{\rho}$ is the tranditional density of states. The authors of \cite{Gibbons:1986xk} found a natural sympletic structure to represent an ensemble of universes. Starting from a generic system with gravitation, $\omega$ may be written in a Darboux basis as
\begin{equation}\omega = d\mathcal{H}\wedge dt + \sum_{i}dp_i\wedge dq^{i},\end{equation}
where $\mathcal{H}$ is the Hamiltonian and $t$ is time. Of course, General Relativity is a highly constrained system \cite{Arnowitt:1962hi}. On such constraint is that $\mathcal{H}$ vanishes. This naturally selects a symplectic structure on the reduced phase space
$$\omega_C = \sum_{i=1}^{n}dp_i\wedge dq^{i}.$$
The natural measure on the constraint hypersurface is then given by $\omega_C^{n-1}\Big|_{\mathcal{H}=0}$. With slight modifications detailed below, this is the measure used by \cite{Gibbons:2006pa}.

Operationally, the measure appends a statistical weight to different trajectories in gravity-scalar phase space. This approach yields a divergent result \cite{Hawking:1987bi}. In \cite{Gibbons:2006pa}, Gibbons and Turok revisited this measure and observed two things. First, as the phase space is three dimensional, with a generic point\footnote{These are not canonical variables, but proxies suitable for this discussion.} $(\phi,\dot{\phi},H)$, familiar results from vector analysis apply. In particular, an analog of ``Ampere's Law'' simplifies the measure to be given implicitly by the normalization of phase space,
$$\int \omega_C^{2}\Big|_{\mathcal{H}=0} = \oint_{S} d\phi\; \frac{1}{a^3}\Big|\frac{d H}{d\phi}\Big|,$$
with 
$$\Big|\frac{d H}{d\phi}\Big| = \sqrt{H^2 - \frac{V}{3}}.$$
In short, one can count all trajectories by integrating over a single ellipse at fixed $H$. 

Second, this measure was regularized by equating physically indistingushable universes. The divergences observed in \cite{Hawking:1987bi} were identified with the ``dilatation symmetry'' of the FLRW metric. The proposed solution was identifying all universes that had been expanded to more than some maximum number e-foldings, $N_{\rm MAX}$. Essentially this observable was ``projectivized'', so the likelihood of e-foldings were computed as ratios $\big<N_{e}\big>/\big<N_{\star}\big>$ with some fiducial $N_{\star}$. Since we are comparing other possibilities to our own, observed universe, such ratios are natural objects to consider.

We now rephrase things in our notation. Since we are on a surface of constant $H$, the variables $(\phi,\phi^{\prime})$ correspond to simple rescaling of relevant plane in phase space. Up to a normalization, $a_0$, the scale factor depends on the number of e-foldings accumulated when each trajectory intersects the ellipse,
$$a = a_0 \exp(3N_e).$$
Therefore, the measure on phase space is given by
\begin{equation}\label{measure}\int \omega_C^{2}\Big|_{\mathcal{H}=0} = \frac{1}{a_0^3}\oint_{S} d\phi\; e^{-3N(\phi)}\sqrt{H^2 - \frac{V}{3}}.\end{equation}

To compute the likelihood of inflation, the inflationary trajectory with $N_{\rm MAX}$ e-foldings was indentified, and a weighed average of $N$ for the other trajectories was applied using the measure in \eqref{measure}. To complete this, one must specify the value of $H$ at which to integrate.
\vskip2ex
The answer depends almost entirely on $H$.

\subsubsection{The Choice of Hypersurface}
For gravity and a neutral scalar field, a CPT invariant measure allows one to consider ensembles with Cauchy data specified at any given time. As the single field inflation is not a chaotic system\footnote{Despite having a three dimensional phase space, a transition to Chaos is not present in this system. This is directly related to the attractor behavior.}, one might simply choose to apply the boundary data at a late time. Specifying the Cauchy data after inflation avoids ambiguities from Planck scale physics, and is closer to our present observations.

Despite being CPT invariant, the effect of the Gibbons-Turok measure depends on what time-like hypersurface you perform the measure calculation. Since we can only observe late-time cosmology, Gibbons and Turok stipulate that we should choose $H$ - and its associated time-like hypersurface - to correspond to the end of inflation.

Recall that the likelihood of inflation, $\big<N_e\big>$, inolves a weighted sum over the e-foldings associated with all trajectories. Each trajectory will be counted at any time-like slicing, but the weight ascribed to each trajectory changes with time. Parameterizing time with $N$ makes this connection manifest. The measure itself, \eqref{measure}, varies as $\exp(-3N)$, so measuring at late times will supress contributions from the inflated trajectories. Indeed, for a given universe $U$ in the ensemble,
$$\mathcal{P}(N_{e}|U)\sim e^{-3N_e},$$
is precisely the result announced in \cite{Gibbons:2006pa}.

This dependence on a choice of time-like hypersurface is a property of the original measure \cite{Gibbons:1986xk}, and has been critiqued before \cite{Hollands:2002xi}. The time dependence has been exploited by Kofman and Linde \cite{Linde:2007fr,kofman} to argue the converse; inflation is almost inevitable if one measure at early times ($H^2\sim M_P^4$). Since all initial conditions have approximately the same statistical weight, a large portion of them yield inflationary trajectories as per the attractor dynamics. Indeed, the effect of late time measurement is to statistically supress almost all of these early time possibilities.

\subsubsection{From Paradox to Resolution via an Ambiguity}
The depedence on a choice of time-like hypersurface yields opposing results, and we've already seen how this tension can be modelled as a conflict between CPT invariance and thermodyamics. But these considerations are only two of many possible time-like hypersurfaces to perform the measurement on. Indeed, varying one's choice of $H$ smoothly interpolates between these possibilities. Without more guidence, this measure still has a glaring ambiguity. 

In the next subsection, we discuss how extending the ensemble of universes to include scalar potential data resolves the interpolation ambiguity into a step function. Better, a choice of $H$ far below the Planck scale can yield arbitrarily large amounts of inflation. This strongly suggests reconsidering a late-time measure surface.

\subsection{Enlarging the Ensemble}

\subsubsection{On Universality}
The ensemble of universes considered so far all contained an arbitrary potential. Indeed, \cite{Gibbons:2006pa} considered large-field inflation; the choice of potential was irrelevant for their considerations. However, the choice of potential is of defining importance for small-field inflation. We have seen how inflection points serve as efficient attractors, effectively scrambling the initial data to give a contribution to inflation whose details depend entirely on the potential \eqref{infKNEE}. To incorporate these details into calculating $\mathcal{P}(N_e|U)$, we must extend the Gibbons Turok measure \eqref{measure} to include scalar potential data.

This is in line with the landscape paradigm. String Theory predicts many possible metastable vacua, each with its own scalar potential. The attractor dynamics described above imply that inflation - however short - is part of a universe settling into such an equilibrium state. Averaging over all such possibilities is a daunting task \cite{Douglas:2003um}, but physical principles guide us into a simple answer suitable for inflation. 

One such principle is \textit{Universality}. That is, there are equivalence classes of theories which give precisely the same physics. The equivalence classes relevant for enlarging the ensemble of universes are those proposed in \cite{Downes:2011gi}. Two theories are considered to belong to the same universality class if their potentials have the same number of parameters near the point of inflation. More precisely, if there exists a neighborhood of the domain where inflation takes places where the potentials have the same germ up to ``versal'' deformations \cite{arnold} parametrized by their couplings. The possible theories are thereby given an ADE classification.

It is this observation that motivates our new definition Multiverse. Starting from the ensemble in \cite{Gibbons:2006pa} we include the space of all possible couplings for an arbitrary smooth potential. Such a large generalization exists because of the domain structure of the parameter space associated to smooth functions. We briefly review the main idea here, but refer the reader to \cite{Downes:2011gi} for details.

Consider a smooth $n$-parameter family of potentials in any number of variables, $V$. Let $\mathcal{M}$ be the space of all parameters, for simplicity assume it to be modelled\footnote{In string theory this should be something like a lattice in $\mathbb{R}^n$.} by $\mathbb{R}^n$. The set of all parameters leading to \textit{degenerate critical points} of $V$ - points where its gradient and Hessian simultaneously vanish - is a (piecewise) smooth hypersurface, $\mathcal{N}$ of codimesion one. In short, only one dimension of parameter space determines whether $V$ is a Morse function (i.e. has no degenerate critical points) or not.

Only in the vicinity of $\mathcal{N}$ is the function $V$ suitable for inflection point inflation. Generically such functions have an expansion of the form \eqref{need}. For multiple fields, one must take care to ensure the inflection point of the inflaton is located at a minimum of the spectator fields. 

This analysis tell us that the likelihood of having an inflection point suitable for inflation depends only on a single dimension in parameter space. For \eqref{need}, this would correspond to parameter $\lambda_1$. As it turns out, this can be related to the number of e-foldings induced by the inflection point. The upshot of all this formalism is the observation that likelihood of e-foldings induced by any inflection point is independent of the choice of $V$. It varies inversely with $\sqrt{\lambda_1}$, as per \eqref{infKNEE}.

Specifically, this means that as $\lambda_1$ approaches zero, (ie, as one approaches a hypersurface of degenerate critical points $\mathcal{N}$ in $\mathcal{M}$), the number of e-foldings vary as $1/\sqrt{\lambda_1}$. After some technicalities derived explicitly below, the likelihood of having $\big<N_e\big>$ e-foldings of \textit{inflection point inflation} is given by,
$$\mathcal{P}(N_e|U) \propto 1/N_{e}^3.$$
Note that $U$ is a universe in the ensemble expanded to include variations of the scalar potential.
\subsubsection{An Example}
\begin{figure}
  \centering
  \subfloat[$\gamma_1$: Curve of constant $N_e$]{\label{fig:cusptwo}\includegraphics[width=0.4\textwidth]{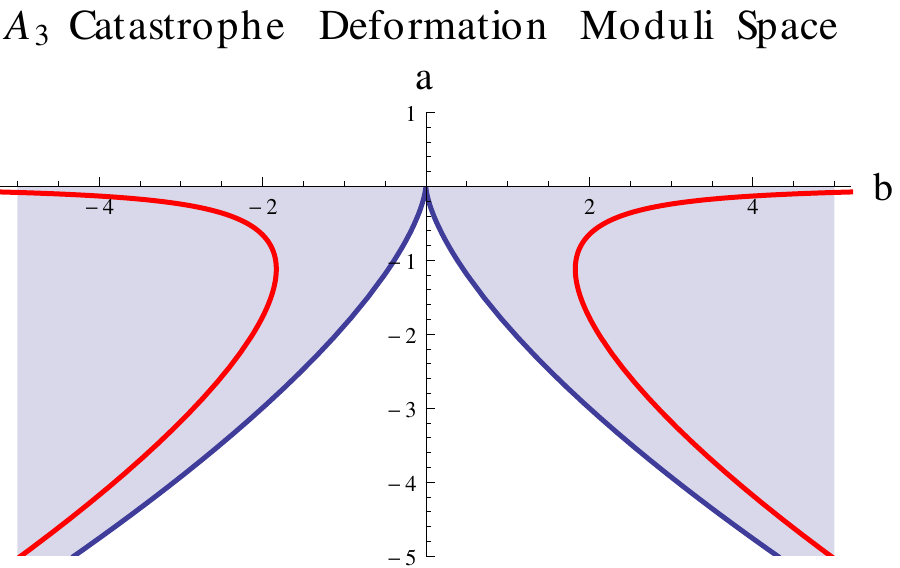}}
  \quad
  \subfloat[The area between $\gamma_1$ and $\gamma_0$]{\label{fig:cuspthree}\includegraphics[width=0.4\textwidth]{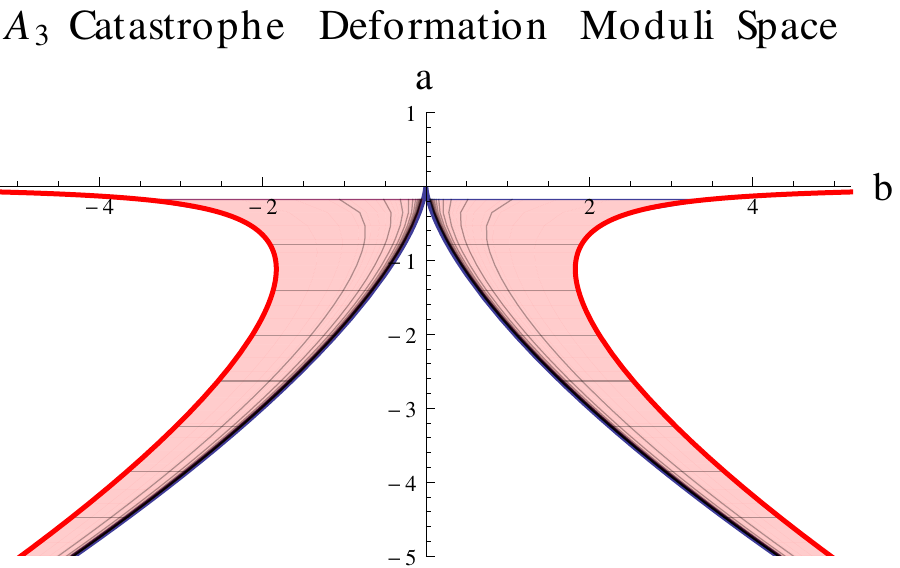}}
  \caption{Excising the relevant portion of parameter space for \textit{at least} $N_e$ e-foldings of inflection point inflation.}
  \label{fig:doublecusp}
\end{figure}

Again we shall demonstrate the main idea with a simple potential \eqref{A4}. We start with the parameters $a$ and $b$ satisfying \eqref{cusp}, so there is an inflection point suitable for inflation. For a more detailed discussion, see \cite{Downes:2011gi}. We also fix the constant term so that the vacuum energy vanishes at the global minimum of $V$. Expanding about the inflection point, $\alpha$, we find,
$$V(x-\alpha) = \frac{1}{4}(x-\alpha)^4 + \alpha (x-\alpha)^3 + \frac{27}{4}\alpha^4.$$
Notice that the position of the inflection point, $x=\alpha$, parametrizes the constant \eqref{cusp}.
\begin{equation}\label{gamma1}a = -3\alpha^2,\quad b = 2\alpha^3.\end{equation}
In the slow roll approximation, this corresponds to an infinite number of e-foldings of inflation. To get a physically reasonable number, we deform the potential slightly. The degenerate critical point has split into two imaginary roots. 
\begin{equation}\label{defocusp}V(x-\alpha) = \frac{1}{4}(x-\alpha)^4 + \alpha (x-\alpha)^3 + \lambda_1 (x-\alpha) + \frac{27}{4}\alpha^4,\end{equation}
The coefficient of the linear term, $\lambda_1$, parametrizes the ``Morsification'' of the potential. It also parametrizes the number of e-foldings, as seen in \eqref{infKNEE}. Fortunately, this all reduces to the simple form,
$$V(x) = \frac{1}{4}x^4 - \frac{a}{2}x^2 + (b+\lambda_1)x + {\rm constant},$$
from which we can read off
$$a = - 3\Big(\frac{b+\lambda_1}{2}\Big)^{2/3}.$$
The area of parameter space associated with least $N_e$ e-foldings of inflation is bounded by two curves, as shown in Fig.~\ref{fig:doublecusp}.

First is $\gamma_0$, given by \eqref{cusp}. The second is $\gamma_1$, which is given by
$$N_e = \frac{\pi}{8\sqrt{3\alpha\lambda_1}}  = \mathrm{constant}.$$
Thus,
$$N_e = \frac{\pi}{8\sqrt{(3a)^{1/2}\lambda_1}}.$$
This implies,
$$\lambda_1 =\frac{\pi^2}{8N_e^2 \sqrt{3a}},$$
which means that $\gamma_1$ is defined by
$$b(a) = -2\left(\frac{a}{3}\right)^{3/2} + \frac{\pi^2}{8N_e^2 \sqrt{3a}}.$$
Therefore the area bounded by the curves $\gamma_0$ and $\gamma_1$ is given by,
$$A_{N_e} = \lim_{A\rightarrow\infty}\int_{0}^{A} \frac{\pi^2}{8N_e^2 \sqrt{3a}}\;da.$$
Thus,
\begin{equation}\label{limiting}A_{N_e} = \lim_{A\rightarrow\infty}\frac{\pi^2}{4N_e^2 \sqrt{3}}\sqrt{A}.\end{equation}
Just like phase space measure, this is a divergent quantity, but the ratio of two areas, for two distinct values of $N_e$ is finite.

Suppose $N_e>N_{\star}$. Then $A_e<A_{\star}$. Therefore, the likelihood of having \textit{at least} $N_e$ e-foldings \textit{relative} to $N_{\star}$ e-foldings is
$$ \frac{A_{N_{e}}}{A_{N_{\star}}} = \left(\frac{N_{\star}}{N_e}\right)^2.$$
To compute the relative likelihood of having precisely $N_e$ e-foldings of inflation to some fiducial amount $N_{\star}$, we take the limit,
$$P(N_e|U) = \lim_{\delta N\rightarrow 0}\frac{1}{\delta N}\Big|\left(\frac{N_{\star}}{N_{e}}\right)^2-\left(\frac{N_{\star}}{N_{e}-\delta N}\right)^2\Big|.$$
Thus,
\begin{equation}\label{knee3}\mathcal{P}(N_e|U)= \left(\frac{N_{\star}}{N_{e}}\right)^3,\end{equation}
for a given universe $U$.

In principle this ratio of infinite areas is ill defined. However, the divergences can be traced to either the limits of extremely large $a$ or $b$, seen as the narrow regions in Fig~\ref{fig:cuspthree}. But $a$ and $b$ parameterize versal deformations of a given Catastrophe germ; they are parameters of some effective field theory obtained by expanding around a particular field configuration. Indeed, \eqref{gamma1} shows that very large $a$ or $b$ corresponds to a very large value of $\alpha$. As $\alpha$ is tied to the scale of inflation, it cannot be arbitrarily large. Since the logarithmic derivative of $V$ enters into the field equations, one might say arbitrarily choosing large $\alpha$ would be allowed so long as we allow an arbitrarily small scaling of the entire potential. As $V$ is quartic in $\phi$ -- and therefore marginal -- this would be technically allowed. Since the allowed initial velocity would also become arbitrarily close to $\phi^{\prime}=\pm\sqrt{6}$, this would bring us back to the original ambiguity posed by \cite{Hawking:1987bi}. However, such a $V$ is presumably only an effective potential, there are likely to be higher order terms, in general such large $\alpha$ would violate naturalness. Therefore, the domain wall moduli like $\alpha$ should be bounded by effective field theory considerations. It is \textit{precisely here} that Planck-scale physics would be invoked to make a complete solution to the Measure ambiguity. Thus, we continue\footnote{It's probably also worth point out that curves of constant e-folding correspond to the classical renormalization group trajectories in parameter space. This is because $\alpha$ is directly correlated with the energy scale. In other words, trajectories won't mix even at large values of $|\alpha|$.} with the simple regularization in \eqref{limiting}.

As discussed above, this simple result is a \textit{general property} of degenerate critical points. It is in remarkable agreement with a recent Monte Carlo analysis \cite{Agarwal:2011wm} performed on D-Brane inflation in the conifold. There up to six fields were considered. Thus, the universality observed in D-Brane inflation and universality discussed in \cite{Itzhaki:2008hs} are both manifestations of the same general features of degenerate critical points. The universality classes of which were discussed in \cite{Downes:2011gi}.

It should be stressed that the details of inflection point inflation are insensitive to noncanonical kinetic terms, and so may be applied in any context (i.e. moduli inflation). The attractor dynamics leading up to inflection point inflation, like the phase transition of Itzaki and Kovetz, can be rather sensitive to such complications; this gives rise to the well-known $\eta$ problem, for example. Noncanonical kinetic terms may be physically important, particularly for primordial non-Gaussianities in multifield models, but we leave them for future investigation.

\subsubsection{Final Thoughts}
Including couplings into the the likelihood analysis of inflation requires the study of inflection points. They rapidly reduce the field velocity, which can generate a substantial amount of inflation, \eqref{infKNEE}. However, inflection points in the potential offer more than an additional mechanism for inflation. By acting as attractor points, they squeeze the trajectories through narrow regions in phase space. This means that the inflation will take place for wide range of initial velocities for a given scalar field VEV. In fact, only a very large field velocity near inflection point generates the overshoot problem.

\begin{figure}[h]
  \centering
    \includegraphics[width=0.9\textwidth]{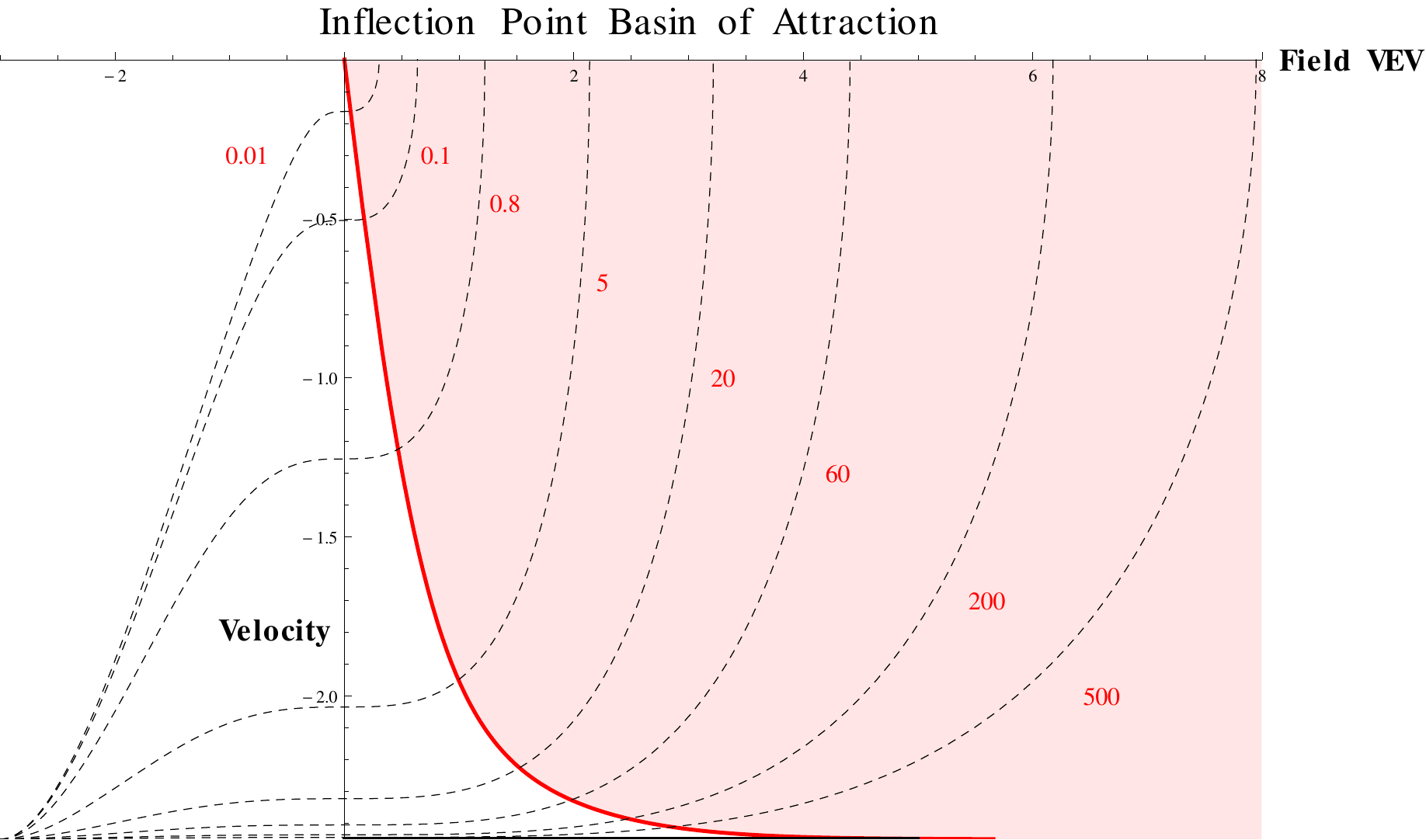}
  \caption{``Phase'' diagram for phase space. Shaded region is the basin of attraction for the inflection point. Dotted lines correspond to different values of Hubble Parameter, in units of $H_{\rm inf} = \frac{9}{4}\alpha^4$ . The feature at the left is the potential's global minimum. See the text for more details}
\label{fig:phase}
\end{figure}

This phenomenon is shown in Fig~\ref{fig:phase}. In this plot of phase space, the shaded region is the basin of attraction for the inflection point; inflection point inflation will definitely occur. Different time-like hypersurfaces (values of $H$) are traced by the dashed lines. From \eqref{hubble}, the Hubble parameter during inflection point inflation is given by $H_{\rm inf}^2 = \frac{9}{4}\alpha^4$. The numbers on the plot corresponds to multiples of $H_{\rm inf}^2$. For instance, the largest contour shown corresponds to a Hubble parameter $H^2 = 500 H^2_{\rm inf}$. We see that larger velocity closer to the inflection point creates overshoot problem just like in the Fig.~\ref{fig:velo}. 

For early times, the inflection point attractor dominates. Note that, in our representation phase space is bounded in the $\phi^{\prime}$ direction. The dark curve representing the inflection point inflation ``phase boundary'' asymptotes to boundary of phase space, $-\sqrt{6}$. Therefore, we can visually see the arguments for CPT invariance versus thermodynamics. In the former case, as time reverses, all trajectories remain in the unshaded region. In the latter, all trajectories have converged at the inflection point.

This dramatic attraction towards the slow-roll trajectory, $\phi^{\prime}\sim -V^{\prime}/V$ can be used to set up an additional period chaotic inflation. Indeed, inflection points can be interpreted as an artificial ``ceiling'' on phase space, in the same way that we typically treat the Planck scale. In this case, however, we \textit{know} that the initial conditions afterwards almost entirely support slow-roll inflation. 

\section{Conclusion}\label{sectionfive}

This work did not aim to make definitive statements about the initial conditions for inflation. We do not propose a mechanism or argument for choosing a time-like hypersurface to measure the Multiverse. Without a solid understanding of the UV dynamics - or at least assurance of a consistent truncation of such a theory - there is no framework for making such predictions. However, these questions pierce the heart of our ignorance. As such, it behooves us to understand these questions with every tool at our disposal. What this work has accomplished, we hope, is the resolution of two the conflicting intuitions of CPT invariance and thermodynamics.

By working within the frame work of \cite{Gibbons:2006pa}, setting aside complications from the UV, we have discussed the role of attractors without invoking entropy production. Our results sharpened the ambiguity of choosing a measure surface by including the effects of inflection point inflation. Is inflation likely to occur, given the usual gravity-scalar sytem? As seen in Fig~\ref{fig:phase}, inflection point inflation resolves the ambiguity to something exponentially close to a yes-no answer. Choosing a value of $H$ below or above the scale of inflection point inflation altered the likelihood of inflation dramatically. Since the scale of inflection point inflation can be considerably lower than chaotic inflation, the influence of attractor dynamics make a late-time measure surface less palatable.



Rather than focusing on either large- or small-field models of inflation, we including the data from the scalar potential in a model independent way; we analyzed the germs of the scalar potential. What emerged was the importance of inflection point inflation in ensemble of universes. 

Put a different way, the Gibbons-Turok exponential suppression associated with inflection point inflation was softened to a power law by allowing the couplings of the scalar potential to vary. 

Given the large number of solutions to string theory and their inherent complexity, a low dimensional jet space of the scalar potential seems a natural fit. Such small-field inflation models are markedly different than their traditionally studied large-field counterparts. They are more akin to critical phenomena. By investigating the Itzaki-Kovetz phase transition further, we have learned more about this relation.

Surely, new phenomena will emerge as one moves to more sophisticated ensembles; it is merely another incarnation of statistical mechanics. There are more areas to explore. Non-canonical kinetic terms have been much studied for similar reasons \cite{Cheung:2007st,Senatore:2010wk}. Such terms are intimately related to the geometry of the four-dimensional effective supergravity models which descend from string theory. They are also known to generate sizable, primordial non-Gaussianities which the Planck satellite is currently investigating. By including them in the ensemble, one might hope to connect measurements of non-Gaussianities to universality classes.

On a more fundamental level, attractors have been a guiding principle in the study of black holes \cite{Ferrara:1995ih} and string theory flux vacua \cite{Kallosh:2005ax}. It would be interesting to see if a suitably augmented attractor mechanism for inflation may further illuminate the initial conditions of our universe. In any case, our study of attractors and universality in the gravity-scalar system has given us a strong reason to believe that inflation - or its embedding into particle physics - will continue to be a predictive framework for understanding the cosmos.

\section*{Acknowledgements}
This work is supported in part by the DOE grant DE-FG02-95ER40917. We would like to thank those at the Mitchell Institute and Cook's Branch where this work was presented. We are particularly indebted to Rouzbeh Allahverdi, Alexey Belyanin, Yu-Chieh Chung Anatoly Dymarsky, Thomas Hertog, and Don Page for insightful feedback. We would also like to thank Jiajun Xu for piquing our interest in \cite{Agarwal:2011wm}. 

\appendix
\section{Probabilities, Likelihoods and all that}\label{sec:prob}

To set notation and make precise the statistical quantities we are interested in, we review the basic terms used in Bayesian statistics, a useful reference is \cite{bayes}. We also establish the general philosphy of this sort of analysis.

We denote the probability of some event, $A$ by $\mathcal{P}(A)$. The conditional probability of an event $A$ \textit{given} an event $B$ is written $\mathcal{P}(A|B)$. Note that $\mathcal{P}(A|B)\neq \mathcal{P}(B|A)$. These distinct probabilities can be related, however, using \textit{Bayes' equation},
$$\mathcal{P}(A|B) = \frac{\mathcal{P}(A)\mathcal{P}(B|A)}{\mathcal{P}(B)}.$$

Typically one is interested in some conditional probability. In cosmology, one speaks loosely of the probability of inflation. Specifically, one means asks: out of some ensemble of universes (or histories), what is the probability of finding one which has undergone sixty e-foldings of inflation. This may be represented by $\mathcal{P}(U|60)$. To answer this question we must first define what we mean by $U$, $\mathcal{P}(U)$, $\mathcal{P}(60)$ and $\mathcal{P}(60|U)$. Then one can apply Bayes' equation. We discuss each of these in turn.

For our purposes, $U$ will be specified by both the initial conditions and the scalar potential. From our discussion in section \ref{sectionthree}, we know that they are not unrelated. Concretely, a choice of $U$ amounts to picking a potential and point in phase space consistent with it. For us consistency means that the chosen point must correspond to an initial vacuum energy below the Planck scale. In this context, the set of all such $U$ is called the \textit{Multiverse}, which we denote by $\mathcal{M}$.

Ignorance of the UV complete theory corresponds to an ignorance of $\mathcal{P}(U)$. It must be put in by hand. This happens often in Bayesian statistics, and such ``subjective'' inputs are called \textit{prior probabilities} or simply priors. Without a large set of universes to perform statistics on, the only educated guess available is the democratic one: all possible universes are equally probable. Therefore, any conclusions drawn from this choice will only be defined up to these ``uninformative priors''. Note that any deviation from these priors -- derived from some vacuum selection rule, say -- would inform us and our calculations of the new physics.

Unlike $\mathcal{P}(U)$, we may say something, albeit tautological, about $\mathcal{P}(60)$. We expand it in terms of conditional probabilities to be
\begin{equation}\mathcal{P}(60) = \sum_{\alpha}\mathcal{P}(U_{\alpha})\mathcal{P}(U_{\alpha}|60).\end{equation}
Here $\alpha$ indexes the possibilities of $U$ described above, and is taken over all $U_{\alpha}$ in $\mathcal{M}$. In \cite{Gibbons:2006pa}, this was defined by \eqref{measure}. Here, in principle, we must also include the contributions for the scalar potential. However, $\mathcal{P}(60)$ will often taken as a normalization. Therefore, the only thing left to define is $\mathcal{P}(60|U)$.

The conditional probability $\mathcal{P}(60|U)$ is something we compute from the dynamics. Since we compute rather than guess, it is a \textit{posterior probability}, and one often calls it a likelihood function. It was what we computed in section \ref{sectionfour}. When we refer to the likelihood of inflation, this is the object of interest.

To be clear, the likelihood of inflation was found to be exponentially suppressed in \cite{Gibbons:2006pa}. By allowing $U$ to contain data from the scalar potential, inflection point potentials were allowed in the ensemble, and this exponential suppression was softened into a power law. While $60^{-3}$ is certainly a tiny number, it is important to keep our priors in mind. Previous computations were done with fixed -- although not necessarily specified -- potentials. Had they chosen a fixed potential with an inflection point, their computation would have resolved the time-like hypersurface ambiguity to something exponentially close to a yes-no answer.

The lesson from this analysis it not the answer to the question posed at the beginning of this appendix. Rather, how a reevaluation of our priors has lead to an enhancement in the likelihood functions $\mathcal{P}(60|U_{\alpha})$ through the emergence of inflection point inflation.

\bibliography{measure}{}
\bibliographystyle{h-physrev}
\end{document}